\documentclass[sigconf, authorversion]{acmart}
\AtBeginDocument{%
  }

\copyrightyear{2025}
\acmYear{2025}
\setcopyright{cc}
\setcctype{by}
\acmConference[ICMI '25]{Proceedings of the 27th International Conference on Multimodal Interaction}{October 13--17, 2025}{Canberra, ACT, Australia}
\acmBooktitle{Proceedings of the 27th International Conference on Multimodal Interaction (ICMI '25), October 13--17, 2025, Canberra, ACT, Australia}
\acmDOI{10.1145/3716553.3750787}
\acmISBN{979-8-4007-1499-3/2025/10}

\begin{document}

\title{Multimodal Behavioral Patterns Analysis with Eye-Tracking and LLM-Based Reasoning}

\author{Dongyang Guo}
\affiliation{%
 \institution{Technical University of Munich}
 \city{Munich}
 \country{Germany}}

\author{Yasmeen Abdrabou}
\affiliation{%
 \institution{Technical University of Munich}
 \city{Munich}
 \country{Germany}}

\author{Enkeleda Thaqi}
\affiliation{%
 \institution{Technical University of Munich}
 \city{Munich}
 \country{Germany}}

\author{Enkelejda Kasneci}
\affiliation{%
 \institution{Technical University of Munich}
 \city{Munich}
 \country{Germany}}

\renewcommand{\shortauthors}{Guo et al.}

\begin{abstract}
Eye-tracking data reveals valuable insights into users’ cognitive states but is difficult to analyze due to its structured, non-linguistic nature. While large language models (LLMs) excel at reasoning over text, they struggle with temporal and numerical data. This paper presents a multimodal human–AI collaborative framework designed to enhance cognitive pattern extraction from eye-tracking signals. The framework includes: (1) a multi-stage pipeline using horizontal and vertical segmentation alongside LLM reasoning to uncover latent gaze patterns; (2) an Expert–Model Co-Scoring Module that integrates expert judgment with LLM output to generate trust scores for behavioral interpretations; and (3) a hybrid anomaly detection module combining LSTM-based temporal modeling with LLM-driven semantic analysis. Our results across several LLMs and prompt strategies show improvements in consistency, interpretability, and performance, with up to 50\% accuracy in difficulty prediction tasks. This approach offers a scalable, interpretable solution for cognitive modeling and has broad potential in adaptive learning, human–computer interaction, and educational analytics.
\end{abstract}




\maketitle

\section{Introduction}

In recent years, large language models (LLMs) have made artificial intelligence (AI) far more accessible through natural language interactions~\cite{mann2020language}. This accessibility allows non-expert users to create complex task workflows by leveraging LLMs' reasoning capabilities. Research has demonstrated that LLMs achieve human-level performance in tasks such as semantic parsing and text generation~\cite{chang2024survey}, and their ability to translate unstructured instructions into executable logic~\cite{gao2023pal} is transforming productivity across domains like education and healthcare. Despite these advancements, applying LLMs to the analysis of structured numerical data—such as time-series physiological signals—remains a significant challenge~\cite{wen2022transformers}. On the other hand, LLMs offer greater flexibility, but still struggle with interpretability, robustness, and reliability in numerical data analysis~\cite{zhao2024explainability}. These limitations highlight the urgent need for collaborative frameworks that combine the computational strengths of AI with human expertise. Such frameworks can enhance trustworthy, data-driven decision-making by integrating contextual judgment with algorithmic insights.

Current research tends to focus on enhancing the autonomous analytical capabilities of LLMs~\cite{huang2022large}. However, two persistent challenges remain. First, AI-generated results often lack transparency, raising critical concerns about their reliability, especially in high-stakes domains like scientific research or industrial optimization~\cite{saeed2023explainable}. Second, while anomaly detection techniques have shown effectiveness in modeling temporal physiological data, they are frequently disconnected from semantic context, limiting their applicability for dynamic, context-aware user guidance~\cite{festag2021semantic}. On the other hand, eye-tracking technology has emerged as a powerful, non-invasive method for collecting high-quality physiological signals~\cite{kasneci2015online,kasneci2024introeyetracking}. It enables the detailed recording of visual behavior such as fixation points, fixation durations, and saccades during cognitive and visual tasks~\cite{skaramagkas2021review,tafaj2013online}. In domains such as education, eye-tracking data helps in analyzing students’ attention distribution and cognitive load during reading or problem-solving tasks~\cite{caldani2020visual,appel2019predicting,castner2018scanpath}, thus supporting the design of more personalized teaching strategies. However, the manual analysis of such high-resolution, large-volume physiological data remains resource-intensive and expertise-dependent.

To address these challenges, we propose a novel eye-tracking data analysis framework powered by large language models (LLMs). The framework enables deep exploration of behavioral patterns in numerical eye-tracking data while supporting robust validation and interpretation through human–AI collaboration. Additionally, we introduce a time-series module for detecting anomalous participant behaviors during experiments. Bridging LLM-driven numerical analysis and cognitive pattern extraction from eye-tracking data, our framework introduces three key components: 1) Multi-Stage Collaborative Mechanism, 2) Expert–AI Co-Evaluation Module, and 3) Anomaly Detection via LSTM and LLM Integration. Our findings show that our collaborative framework improves the interoperability of eye-tracking data analysis compared to traditional AI-only or expert-only methods. The integration of expert judgment and AI reasoning enhances detection and explanation of user states, making the analysis more transparent and actionable.

\section{Related Work}
Our work builds on two areas: 1) Application of LLMs, and 2) Eye-tracking data analysis. The following section reviews the literature in these areas to contextualize our framework.

\subsection{Application of Large Language Models on Numerical Data}
Before the emergence of large language models (LLMs), researchers primarily focused on manual analysis~\cite{ding2024data}, scripts~\cite{hacking2023comparing}, and leveraging neural networks to process and analyze data, such as generating and applying tabular data using machine learning techniques~\cite{fang2024large,badaro2023transformers,mitchell1999machine}. The advent of LLMs has broadened the scope of numerical data processing and analysis. Consequently, researchers have begun to explore key challenges LLMs face when handling numerical data, including multimodal data integration~\cite{huang2023chatgpt, he2024exploring}, interpretability~\cite{ben2024lvlm, singh2023augmenting, yang2023towards}, data quality~\cite{yu2024makes, lee2023making}, and data security~\cite{yao2024survey, he2023large, brown2022does}.
Despite the remarkable success of LLMs in natural language processing and domain-specific tasks~\cite{chang2024survey,KASNECI2023102274}, their performance in processing structured numerical data remains limited. To address this, Wei et al.~\cite{wei2022chain} proposed an in-context reasoning approach that converts numerical problems into interpretable intermediate steps through structured reasoning processes. This method enhances the model’s sensitivity to numerical reasoning and brings its performance closer to systematic human-like thinking, significantly improving numerical accuracy. 


Further studies, such as the one done by Li et al.~\cite{li2024common}, demonstrated that optimizing data strategies—rather than increasing model size, can substantially improve LLMs' capabilities in solving numerical problems. This finding was validated through LLM-generated mathematical datasets. Despite the significant efforts from the research community, the performance of large language models (LLMs) in processing structured numerical data still leaves substantial room for improvement. Moreover, the exploration of LLMs for pattern detection across numerical datasets remains relatively underdeveloped. In a recent study, Guo et al.~\cite{guo2024towards} applied LLMs to traffic flow prediction tasks for the first time. By transforming multimodal traffic data into natural language descriptions and leveraging the reasoning capabilities of LLMs, the approach not only enabled accurate predictions but also effectively addressed the lack of interpretability often associated with deep learning models. To further tackle the limitations of LLMs in complex geo-spatial tasks. Furthermore, Chen et al.~\cite{chen2024llm}, proposed an interactive framework that integrates a code interpreter, static analysis, retrieval-augmented generation (RAG), and Monte Carlo Tree Search (MCTS). This framework incorporates external geo-spatial knowledge (e.g., library documentation and domain-specific methodologies) via RAG and uses MCTS to explore optimal analytical pathways. As a result, it significantly mitigates the hallucination problem in LLMs, enhancing both the reliability and interpretability of the model in multimodal reasoning contexts.

\subsection{Analyzing eye tracking data patterns}
Eye-tracking technology has become a vital tool for behavioral analysis in various fields such as education~\cite{wang2020does,appel2021cross,kasneci2022your,buhler2024task}, psychology~\cite{hessels2019eye}, and marketing~\cite{ke2024using}. Eye tracking enables precise recording and analysis of eye movement patterns as a key method for assessing visual attention and inferring cognitive processes~\cite{skaramagkas2021review}. This allows researchers to reveal how individuals allocate visual attention during specific tasks, thereby providing insights into their cognitive processes and decision-making behaviors. 


 Hence, Saiz et al.~\cite{saiz2020lifelong}, integrated eye-tracking technology with data mining techniques to analyze participants’ visual attention distribution and cognitive mechanisms during task execution. By identifying effective versus ineffective learning patterns, the study provided data-driven support and strategic guidance for personalized education. 
 

 Vsola et al.~\cite{vsola2024ai} was the first to introduce consumer behavior prediction algorithms from neuroscience into the educational domain by integrating eye-tracking data with electroencephalogram (EEG) signals. The study developed a dynamic cognitive load assessment model tailored for online learning environments. This model captures students’ attention distribution in real time as they watch instructional videos and employs statistical analysis using the R programming language to quantitatively reveal the relationship between cognitive load and instructional content complexity. 
 
 Expanding on this work, Aksu et al.~\cite{aksu2024mental} achieved the first synchronized integration of EEG and eye-tracking data in cognitive workload assessment. The study used the characteristics of two physiological signals, electroencephalogram (EEG) and eye tracking, to capture real-time brain activity and characterize visual attention distribution, respectively. They revealed the dynamic relationship between task complexity and cognitive load, and proposed a standardized process for multimodal physiological data analysis. This provides a replicable technical foundation for future research in this field.

Although eye-tracking technology has been widely applied in behavioral analysis, existing research has primarily focused on the extraction of explicit features—such as fixation hotspots and saccade paths—and their correlation with task performance. However, the exploration of deeper semantic patterns embedded in eye movement sequences, such as cognitive states and decision-making logic, remains relatively limited. In particular, there is a lack of systematic exploration into context-aware semantic reasoning and pattern recognition on eye-tracking data in combination with large language models (LLMs). 

Hence, in this work, we contribute a novel human–AI collaborative framework for eye-tracking data analysis that (1) introduces a multi-stage analysis pipeline for uncovering latent patterns, (2) establishes a co-evaluation model for trust scoring between experts and LLMs, and (3) enables real-time anomaly detection through LSTM-LLM integration. Our approach leverages the effectiveness of the framework and provides a scalable solution for intelligent interaction design in areas such as education, psychological assessment, and adaptive user interfaces.


\begin{figure*}[t]
    \centering
    \includegraphics[width=\linewidth]{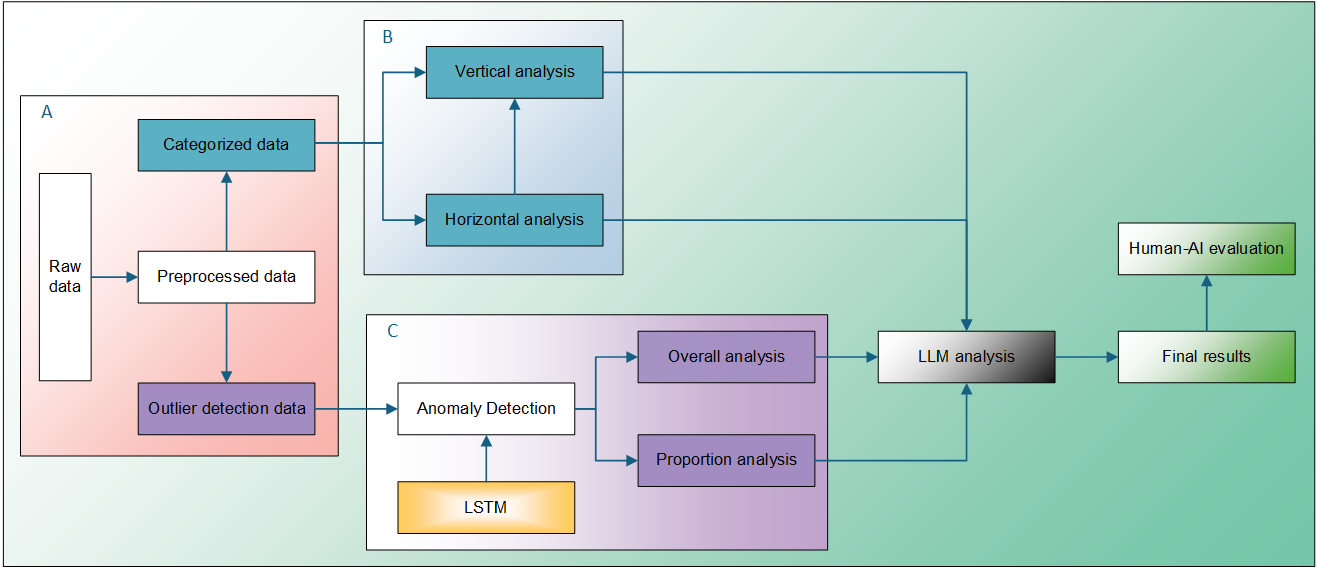}
    \caption{Multi-module integrated process design for attention shift and cognitive load detection. a) represents data processing. b) represents the two-dimensional splitting process. c) represents the abnormal behavior detection process.}
    \label{fig:1}
\end{figure*}

\section{Multimodal Framework}

We introduce a human–AI collaborative framework for eye-tracking data analysis that integrates large language model (LLM) reasoning, expert input, and temporal modeling. We aim to not only support conventional analytical tasks efficiently but also to uncover latent cognitive patterns that traditional methods often overlook. By applying this framework to educational assessment scenarios, we aim to provide methodological support for cognitive science and human–computer interaction research for leveraging eye-tracking data across diverse domains. The framework is structured around three core modules:

\texttt{Multi-Stage Collaborative Mechanism}:
A structured, multi-phase analysis pipeline that segments and processes eye-tracking data both horizontally and vertically. This mechanism integrates automated reasoning with expert input to uncover latent attention and behavior patterns, enhancing the robustness and interpretability of results.

\texttt{Expert–AI Co-Evaluation Module}:
A co-scoring system in which LLM-generated analyses are evaluated in collaboration with domain expert. The module produces confidence scores that quantify the credibility of behavioral patterns, offering a more reliable foundation for downstream decision-making.

\texttt{Anomaly Detection via LSTM and LLM Integration}:
A hybrid component combining LSTM-based temporal modeling with LLM-driven semantic interpretation. This module enables real-time detection of anomalies such as attention shifts, cognitive load fluctuations, and learning difficulties, providing actionable insights for optimizing experimental engagement.

This modular architecture ensures interpretability, adaptability, and improved reliability in analyzing eye-tracking data during cognitively demanding task.

\subsection{Dataset}
To develop our framework, we utilized an existing dataset published by Jang et al.~\cite{jang2024exploring}, which captures the collaborative dynamics of pair programming through eye-tracking data. The study involved 19 participants (9 students and 10 experts), recorded using a Tobii Pro Fusion device~\footnote{Tobii Pro Fusion: \url{https://connect.tobii.com/s/fusion-get-started?language=en_US}} at 250Hz, across three pairing conditions (student–student, expert–expert, student–expert) and two roles (driver and navigator). The experiment aimed to examine how participants manage code exploration and communication during collaborative programming. 



This eye-tracking dataset provides objective data from participants as they complete programming tasks of varying difficulty. It reveals how differences in programming expertise influence collaboration strategies and how task complexity interacts with behavioral patterns. Importantly, the dataset adopts a tabular structure that aligns seamlessly with our proposed framework. Each row or column contains structured fields such as fixation\_number, saccade\_number, and saccade\_duration. Through our method, we uncovered subtle gaze behaviors not explicitly analyzed in the original study—such as role-specific shifts toward error regions. These new findings add an interpretable and scalable layer of behavioral analysis to the original dataset, bridging numerical eye-tracking signals with semantic reasoning. This application demonstrates the broader potential of our framework for cognitive modeling in educational and human–computer interaction contexts.

\subsection{Workflow}

Figure~\ref{fig:1} illustrates the overall workflow of our framework. As shown in region A of Figure~\ref{fig:1}. We began by cleaning the raw eye-tracking data to eliminate noise, irrelevant entries, and missing values. The cleaned dataset was then duplicated: one version was annotated by assigning each fixation point to predefined Areas of Interest (AOIs) based on the experimental design, while the other remained unclassified to preserve the original structure

As illustrated in region B of Figure~\ref{fig:1}, the annotated dataset was first processed by the horizontal analysis module, which splits the data across the temporal axis to uncover relationships between features at specific time points. The results were fed into a large language model (LLM) to generate relevant insights and preliminary behavioral patterns. In parallel, the data were passed through the vertical analysis module, which performs longitudinal splitting to identify feature evolution over time. Outputs from both modules were merged and reanalyzed by the LLM to extract deeper semantic patterns and structural dependencies.

To detect anomalies in participants' eye-tracking data, we introduced Long Short-Term Memory (LSTM) networks, as illustrated in area C of Figure~\ref{fig:1}. Specifically, we trained an LSTM time-series model based on expert-level participants' eye-tracking data, and quantified deviations between the eye movement patterns of student-level participants and expert behaviors using reconstruction error. This enables task-driven anomaly detection. Subsequently, we input the distribution features of anomalies within each AOI into the LLM, analyzing them from the perspectives of group differences and individual stability. The system revealed significant differences in cognitive behavior patterns between student-level and expert-level participants.

Finally, we conducted a validation phase to assess the reliability of the LLM-generated behavioral patterns. To reduce uncertainty during the generation process, we ran the LLM 10 times for each experimental phase and combined the results into a set of behavioral patterns. We then deduplicated the behavioral patterns within the set to ensure diversity and representativeness. From this set, we randomly sampled and validated the behavioral patterns using a proportional sampling method. Specifically, we assessed each behavioral pattern through a dual evaluation process, combining expert judgment with theoretical support from relevant literature, and calculated the Cohen's Kappa coefficient \cite{cohen1960cofficient} to measure the consistency between the judgments. A Kappa coefficient above 0.6 indicated high consistency and reliability in the behavioral patterns.

\subsection{Data Analysis Pipeline}

\begin{figure}[t]
    \includegraphics[width=1\linewidth]{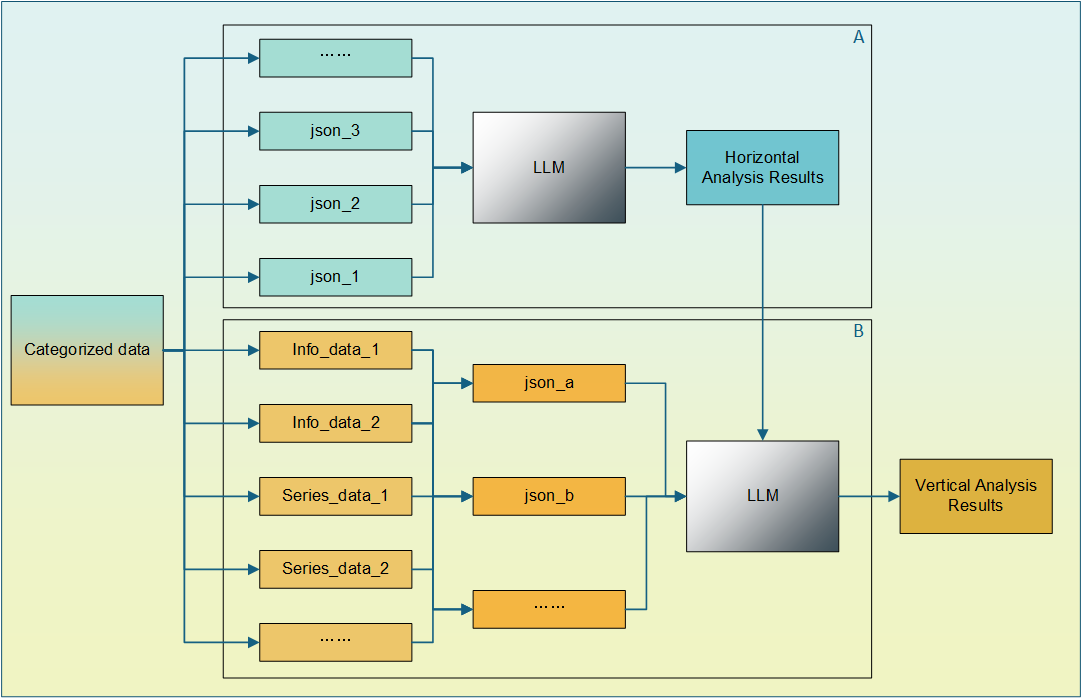}
    \caption{LLM-based 2D Data Analysis Flowchart: (A) Blue path – horizontal split (rows to JSON for LLM); (B) Orange path – vertical split (ID-column pairs to JSON), analyzed with horizontal results.}
    \label{fig:2}
\end{figure}

Figure~\ref{fig:2} shows our proposed LLM-based two-dimensional data analysis pipeline, which mainly consists of two stages:

\begin{itemize}
\item {\texttt{Horizontal analysis module (blue path)}}: We first split the original structured data into rows, each row is regarded as an independent unit, and converted into JSON format. After that, the data is input into the large language model (LLM) for analysis. This process aimed to uncover horizontal associations between fields within the same data instance, yielding a set of potential behavioral patterns.

\item {\texttt{Vertical analysis module (orange path)}}: We regard each column of data as the basic unit of analysis. Specifically, we first extracted columns with identifiable properties (ID attributes) and then paired them with other numerical columns to form tuples or JSON files. Based on this, we input the combined data along with the behavioral patterns generated in the horizontal analysis phase into the LLM to further uncover latent patterns and structural features from the vertical dimension.
\end{itemize}

\vspace{-10pt} 

\subsection{Outlier detection}
We developed an anomaly detection module based on Long Short-Term Memory (LSTM) networks. This module uses the eye-tracking data of expert-level participants as the training set. The data has been preprocessed and cleaned. A time-series model is then built to learn their typical fixation patterns during task execution.. Subsequently, this model detects anomalies in the eye-tracking data of student-level participants, identifying deviations in their fixation behaviors relative to the expert paradigm. After detection, we input the distribution of the anomalies across different Areas of Interest (AOIs) into the large language model (LLM). The LLM analyzes the latent patterns in the anomaly distribution. It then uncovers cognitive differences and behavioral characteristics within the student group during task execution. Based on this analysis, it generates corresponding behavioral patterns.

\section{Evaluation Methods and Metrics}

In order to analyse our data, we implemented four analysis steps, \textit{Pattern Mining} for mining the potential information behind eye movement data, \textit{Consistency of behavioral patterns} for verifying whether the information we extract is credible, \textit{Difficulty Prediction} for verifying whether our framework has accurate teaching capabilities, and finally, \textit{Anomaly Detection} for verifying whether our framework can detect students' abnormal learning behaviors. Below, we explain each one in depth.

\subsection{Pattern Mining}
We selected three mainstream large language models—ChatGPT-4o\footnote{ChatGPT-4o: \url{https://openai.com/index/hello-gpt-4o/}}, ChatGPT-o1\footnote{ChatGPT o1: \url{https://openai.com/o1/}}, and Deepseek-R1\footnote{Deepseek-R1: \url{https://www.deepseek.com/}} to validate the proposed analysis framework. We designed three levels of prompts to meet different input requirements: 

\begin{itemize}
    \item \textbf{Detailed prompt}, which explicitly states the data background and highlights the key information to focus on.
    \item \textbf{Semi-detailed prompt}, providing the complete data background without specifying the analytical focus
    \item \textbf{Brief prompt}, which contains no data background or analytical guidance. 
\end{itemize}

Details of the three prompts are in the Pattern Mining section of the Appendix. We then combined the three types of prompts with the three analysis modules for validation: the horizontal analysis module (H), the vertical analysis module (V), and the horizontal + vertical combined module (H+V), generating multiple sets of preliminary behavioral patterns. To reduce the time and financial costs associated with manual evaluation, we further used Deepseek-R1 to perform inductive analysis on all behavioral pattern sets. To ensure consistency, we kept the prompt type and analysis module fixed while comparing the behavioral patterns generated by different models. We label the behavior patterns that appear in multiple models as high frequency, and the behavior patterns that appear in only one model as low frequency. Because high-frequency behavioral patterns are highly representative, they can represent the true views of LLM. To make the selected behavioral patterns representative and save computing resources, we choose to extract 30\% of high-frequency behavioral patterns and put them into the new behavioral patterns set. Similarly, low-frequency behavioral patterns may be LLM’s unique views, but they may also be due to their hallucination. These behavioral patterns are often not representative, so we select 10\% of low-frequency behavioral patterns and put them into the new behavioral patterns set.

\subsection{Consistency of Behavioral Patterns}
To assess the consistency between expert manual evaluation and the automatic analysis performed by the large language model, we employed Cohen's Kappa coefficient ($\kappa$). The Kappa coefficient effectively controls for bias introduced by chance agreement, providing a more scientifically accurate reflection of the true level of agreement between expert ratings. The calculation of this coefficient relies on a dual validation process for each behavioral pattern. In our framework, we first invite a domain expert, who has a solid understanding of the data background, to perform the first round of evaluation on the newly generated behavioral pattern set. Subsequently, we conduct a second round of validation using literature-based verification: for each behavior pattern, we input it into the large language model and ask it to recommend 5 related papers. We then performed a comparative analysis of the five related research papers to determine its effectiveness and rationality in the context of existing research.
To save time and costs associated with manual literature review, we introduce a large language model to assist in the literature reading and content rating tasks, thereby enhancing overall efficiency while maintaining the quality of validation. Finally, based on the dual results from expert ratings and literature support, we perform a weighted scoring to comprehensively assess the correctness of each behavioral pattern. We outline the evaluation dimensions in Table~\ref{tab:1}, and we used equation ~\ref{eq:1} and ~\ref{eq:f}:

\begin{table}[b]
  \caption{Scoring Criteria (C = Citation Count, R = Journal Ranking, S = Behavioral Pattern Stance)}
  \label{tab:1}
  \begin{tabular}{ccl}
    \toprule
    Dimension&Weight Allocation\\
    \midrule
      C  & 1st: 5 pts, 2nd: 4 pts, 3rd: 3 pts, 4th: 2 pts, 5th: 1 pt\\
      R & JCR Q1: 4 pts, Q2: 3 pts, Q3: 2 pts, Q4: 1 pt \\
      S & Support: +1, Oppose: -1, Neutral: 0 \\
  \bottomrule
\end{tabular}
\end{table}
\vspace{-.35cm}

\begin{equation}
F_i = (C + R) \cdot S, \quad i \in \{1,2,3,4,5\}
\label{eq:1}
\end{equation}

\begin{equation}
F = \sum_{i=1}^{5} F_i
\label{eq:f}
\end{equation}

Based on the final weighted score \(F\), we assess behavioral pattern validity: if \(F>0\), the pattern is valid; if \(F<0\), it is invalid. After this, we calculate Cohen's Kappa to measure consistency between expert evaluation and literature validation, as shown in equation~\ref{eq:2}.

\begin{equation}
\kappa = \frac{p_o - p_e}{1 - p_e}
\label{eq:2}
\end{equation}

Here,  $p_o$ represents the observed agreement proportion (i.e., the frequency of agreement between the two judgments), and  $p_e$  represents the expected agreement under the assumption of random judgments by the evaluators. If $k>0.6$, it indicates that the behavioral pattern has high consistency and a high level of reliability. Following this method, we conducted a Kappa coefficient analysis on two sets of behavioral patterns: one set is generated by the LLM directly based on the raw data, and the other is a composite behavioral pattern set generated after combining the three modules (H, V, H+V) with the three types of prompts.

\subsection{Question Difficulty Prediction}
To further explore the potential of LLMs in educational assessment scenarios, we focus on their ability to predict question difficulty levels in the context of programming task design. Based on cognitive load theory, we constructed an experimental framework to evaluate this capability. The dataset consists of programming questions that have been pre-categorized into three difficulty levels: easy, medium, and hard, with four questions in each category. To ensure objectivity in the experiment, we first anonymized the questions by removing all explicit or implicit indicators related to their original difficulty labels. Next, to evaluate the impact of different prompting strategies on the model's prediction performance, we designed two types of prompts: the total prompt, which explicitly states that there are 12 questions evenly distributed across three difficulty levels (easy, medium, and hard), and the none prompt, which provides no information about the number of questions or the distribution of difficulty levels. We then input the anonymized questions into the LLMs and observed how well they inferred and classified the difficulty levels under these different prompting conditions. To further improve prediction accuracy and investigate the benefits of structured analysis for this task, we integrated our analytical modules: Horizontal Analysis (H) and Vertical Analysis (V). Three experimental settings were tested: (1) using the H module only; (2) using the V module only; and (3) using both H and V modules simultaneously (H+V). Experiments were conducted across three large language model platforms: ChatGPT-4o, ChatGPT-o1, and Deepseek-R1. 
Considering the potential variability in output from large language models, each experiment setting was repeated 5 times, and the final prediction accuracy was calculated as the average across runs to ensure robustness and representativeness. We use prompts from the Question Difficulty Prediction Prompts section of the Appendix in this part.

\begin{figure}[t]
    \centering
    \includegraphics[width=1\linewidth]{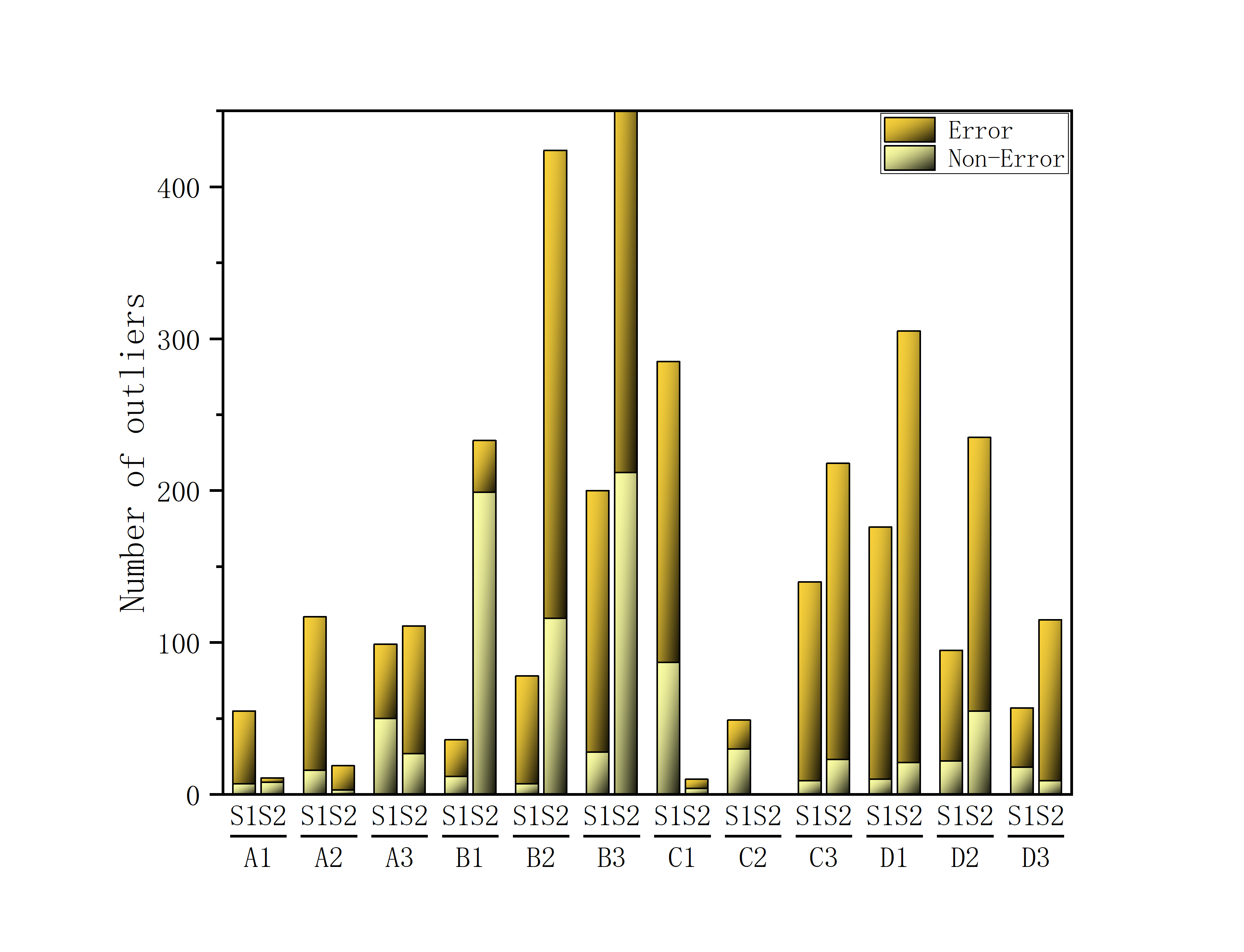}  
    \caption{Outlier distribution: A1–D3 (questions), S1–S2 (students), Error (problem area), Non-Error (question stem).}
    \label{fig:3}
\end{figure}
\vspace{-.35cm}

\subsection{Anomaly Behavior Detection}
Our analytical framework also incorporates an anomaly detection–based pipeline. It is designed to support the interpretation of behavioral differences across participant groups during task execution. We first developed an anomaly detection module based on a Long Short-Term Memory (LSTM) network. This module was trained on preprocessed eye-tracking data from expert participants to learn prototypical gaze patterns exhibited during task performance. We subsequently applied the trained model to the eye-tracking data of student participants to detect deviations from expert-like patterns of visual attention. 

To provide a more intuitive understanding of the detection outcomes, we visualized the anomaly distributions throughout task execution. Figure~\ref{fig:3}, highlights the anomaly distribution patterns of two student participants across the full task sequence (A1–D3), capturing potential shifts in attention and fluctuations in cognitive engagement at different task stages. Beyond visualization, we fed the anomaly distributions within Areas of Interest (AOIs) into a large language model (LLM)-powered analysis system and established a dual-dimensional semantic mining framework. The prompts we use are in the Anomaly Behavior Prompts section of Appendix. On one hand, we conducted horizontal inter-group comparisons to reveal structural differences in anomalous behaviors between student and expert groups. On the other hand, we performed vertical intra-individual pattern analysis to trace the evolution of cognitive strategies and behavioral trajectories within individual students throughout the task.
This anomaly-driven pipeline enhances the semantic interpretation of detection results. It also introduces a new way to use LLMs in educational assessment and cognitive modeling.

\section{Results}
This section presents experimental results in anomaly detection, behavioral pattern consistency, and difficulty prediction, supporting the effectiveness and robustness of our human–AI collaborative framework across various models and configurations.

\subsection{Anomaly Behavior Detection}

Our anomaly detection pipeline combined LSTM-based temporal modeling with LLM-driven semantic analysis to identify deviations in student eye-tracking patterns compared to expert baselines. The results revealed meaningful differences in cognitive behaviors:
\begin{itemize}
    \item  S1 showed a relatively stable attention pattern, with anomalies evenly distributed and a lower error rate on simple tasks (15.5\%). This suggests that S1's performance aligns well with expected difficulty levels and expert-like engagement on easier questions.
   
    \item S2 exhibited a concentration of anomalies on both difficult and certain simple questions (e.g., B3), suggesting issues such as distracted attention, ineffective strategies, or weak foundational knowledge.
    
    \item Recommendations: S2 should focus on questions B1 and D2 and review errors on B3. S1 may benefit from closer analysis of performance on C1 (medium) and A3 (difficult).
   
    \item Task Design Insight: Question C2 may contain flaws, indicated by a double-zero anomaly, warranting redesign.
\end{itemize}

The ability of the framework to generate both personalized feedback (e.g., S1/S2 suggestions) and systemic critique (e.g., C2 redesign) highlights its practical value in adaptive learning systems. We put more examples in the Anomalous Behavioral Patterns section of the Appendix.

\subsection{Consistency of Behavioral Patterns}
Through our method, we obtain a large number of behavioral patterns, which we show in Behavioral Patterns section of Appendix. We evaluated the performance of the proposed multi-stage pipeline in the task of behavioral pattern mining. The final trust scores were generated using our expert-model collaborative scoring module, and the experimental results are shown in Table~\ref{tab:2}. The Directly setting represents the baseline condition where no segmentation modules were applied. Under this setting, Deepseek-R1 achieved the highest trust score (0.715), followed by ChatGPT-o1 (0.647) and ChatGPT-4o (0.524). These results represent the lower bound of model performance without structural processing or enhanced prompting. In the total (fully detailed prompts) condition, the simultaneous application of both horizontal (h) and vertical (v) modules significantly improved trust scores across all models: 4o reached 0.75, while both o1 and r1 achieved perfect scores of 1.00. This demonstrates that when provided with sufficient structural cues and task instructions, the models can achieve ideal levels of behavioral interpretation. In the half (partially detailed prompt) condition, the h+v combination maintained strong performance, with o1 and r1 still scoring 1.00, and 4o showing a slight decrease to 0.687. It is worth noting that even in the absence of additional prompting (none condition), the h+v module combination still yielded relatively high trust scores—for example, 0.818 for o1—demonstrating the robustness of the structured analysis approach under weakly guided scenarios.




\begin{table}[t]
  \caption{Consistency comparison: Directly (no module), h (horizontal), v (vertical); total (detailed), half (semi-detailed), none (brief); 4o (ChatGPT-4o), o1 (ChatGPT-o1), r1 (Deepseek-R1).}
  \label{tab:2}
  \begin{tabular}{cccc}  
    \toprule
     & 4o & o1 & r1 \\  
    \midrule
    Directly & \textbf{0.524} & \textbf{0.647} & \textbf{0.715} \\
    h+v(total) & \textbf{0.75} &\textbf{1} & \textbf{1} \\
    h(total) & 0.696 & 0.764 & 0.813 \\
    v(total) & 0.727 & 0.818 & 0.799 \\
    h+v(half) & \textbf{0.687} & \textbf{1} &\textbf{1} \\
    h(half) & 0.684 & 0.764 & 0.727 \\
    v(half) & 0.671 & 0.788 & 0.843 \\
    h+v(none) & \textbf{0.681} &\textbf{0.818}  & \textbf{0.771} \\
    h(none) & 0.581 & 0.711 & 0.754 \\
    v(none) & 0.541 & 0.792 & 0.722 \\
  \bottomrule
  \end{tabular}
\end{table}




\subsection{Question Difficulty Prediction}

We evaluated the performance of the proposed multi-stage pipeline in the task of question difficulty prediction. The final prediction accuracies are shown in Table~\ref{tab:3}.

Under the Directly(total) setting—where full prompts were provided but no structural analysis modules were used—only ChatGPT-o1 (0.333) and Deepseek-R1 (0.317) produced results, both with relatively low performance. This indicates that without structural guidance, large language models (LLMs) still exhibit limited accuracy in numerical reasoning tasks. With the introduction of both horizontal and vertical analysis modules, the h+v(total) setting led to notable improvements across all three models: 4o reached 0.483, o1 achieved 0.467, and r1 attained 0.50. These results suggest that under fully prompted conditions, structural analysis enables LLMs to better comprehend numerical data and make more accurate difficulty predictions. Under the simplified none prompt condition, the h+v combination still demonstrated a degree of robustness: 4o and o1 both reached 0.417, while r1 achieved 0.40. Although overall performance decreased compared to the fully prompted setting, it remained higher than the baseline. Notably, some models failed to produce valid outputs under specific settings (e.g., o1 under v(none)), possibly due to limited adaptability to minimally guided structural inputs.

We proposed and evaluated a human–AI collaborative framework that combines structured eye-tracking data, expert feedback, and LLM reasoning to extract cognitive patterns. Our modular analysis approach improves the reliability, consistency, and interpretability of LLM-generated insights across tasks like anomaly detection, pattern analysis, and difficulty prediction.




\begin{table}
  \caption{Difficulty prediction accuracy: Directly (no module), h (horizontal), v (vertical); total (detailed prompt), half (semi-detailed), none (brief); 4o (ChatGPT-4o), o1 (ChatGPT-o1), r1 (Deepseek-R1).}
  \label{tab:3}
  \begin{tabular}{cccc}  
    \toprule
     & 4o & o1 & r1 \\  
    \midrule
    Directly(total) & \textbf{NA} & \textbf{0.333} &\textbf{0.317}  \\
    h+v(total) &\textbf{0.483}  &\textbf{0.467}  &\textbf{0.50}  \\
    h(total) & 0.433 & 0.450 & 0.450 \\
    v(total) & 0.30 & 0.367 & 0.35 \\
    h+v(none) & \textbf{0.417} & \textbf{0.417} & \textbf{0.40 }\\
    h(none) & 0.333 & 0.35 & 0.333 \\
    v(none) & 0.283 & NA & 0.317 \\
  \bottomrule
  \end{tabular}
\end{table}

\section{Discussion}
This section reflects on the key implications of our findings and considers the ethical and safety dimensions of applying such systems in educational and cognitive contexts.
\vspace{-.35cm}
\subsection{Human-AI Collaborative Framework}
Our findings confirm that combining human expertise with LLM reasoning via structured analysis modules provides measurable improvements in both consistency and accuracy. The joint use of horizontal (temporal) and vertical (semantic) segmentation consistently outperformed either module in isolation. This suggests that LLMs, when appropriately guided, are capable of engaging with complex numerical and behavioral data, extracting latent cognitive patterns that might otherwise remain obscured.

Notably, our framework proved robust even in low-context settings. When detailed prompt information was unavailable, the modular structure still preserved model performance and consistency. This finding is particularly relevant for real-world deployments in educational environments where data and metadata are often incomplete or inconsistently labeled~\cite{sarah2004quality}. In the anomaly detection task, our system not only highlighted behavioral inconsistencies among students but also identified potential flaws in question design, such as the double-zero anomaly observed in Question C2. This dual diagnostic capacity—focusing on both learner and content—demonstrates the potential of our framework to support both adaptive feedback systems and instructional design processes.

Furthermore, the expert–AI collaborative scoring mechanism has multiple applications. It can serve as a quality control layer, helping researchers and system designers filter and validate LLM outputs. Additionally, the trust scores derived from expert-model alignment can inform personalized feedback systems and adaptive learning pathways, improving user trust and engagement. These scores also provide direct feedback for refining prompt strategies and adjusting model configurations, enabling a closed-loop system for ongoing human–AI co-optimization. Hence, future work should focus on automating the expert feedback integration process to reduce manual annotations and enhance scalability. Furthermore, we aim to expand the framework to support multimodal analysis by incorporating additional data streams such as audio, speech, or physiological signals (e.g., EEG), to enhance the understanding of user cognitive states and increase the applicability of the framework across different domains.

\subsection{Implications for Educational Assessment and Personalization}
By identifying differences in cognitive behavior between students and experts, the framework enables a deeper understanding of learning strategies, attention patterns, and performance shortcomings. These insights could inform personalized tutoring systems that adapt to an individual learner’s cognitive state, task difficulty preferences, or attention fluctuations. Furthermore, the ability of the LLM to predict question difficulty, especially when supported by structured analysis, indicates its potential to assist educators in balancing assessments and scaffolding learning tasks. In contexts where human resources are limited, this could contribute to more equitable, data-driven educational experiences.
\vspace{-.35cm}

\subsection{Generalizability Across Different Domains}
Although this paper is validated using data from an educational context, specifically in programming learning tasks, the proposed analytical framework is designed with strong cross-domain adaptability. The core components—multi-stage gaze pattern mining, expert-AI collaborative evaluation, and a hybrid anomaly detection module that combines temporal modeling with semantic reasoning—exhibit high generalizability. This approach is not only suitable for educational feedback and cognitive monitoring but also holds potential for broader applications such as medical diagnostics\cite{wright2011method}, user interface design\cite{bakke2011spreadsheet}, and workplace training\cite{giacumo2016emerging}, where understanding user attention and cognitive states is critical. Due to its compatibility with various language models and prompt strategies, the system can flexibly accommodate diverse task requirements and user behaviors. Future research could extend evaluation to a wider range of data types and task environments to further assess its robustness and transferability, thereby advancing the broader application of cognitive modeling techniques in multimodal human-computer interaction.

\vspace{-.25cm}
\subsection{Challenges of LLMs with Structured Non-Text Data}
This paper reveals significant limitations of large language models (LLMs) when processing structured non-text data such as eye tracking data. LLMs lack direct parsing capabilities for numerical data with low semantic density. For example, when eye tracking data is directly fed into the model (such as ChatGPT-4o), it cannot even effectively identify information (classification accuracy is NA). This indicates that the model has difficulty in autonomously establishing semantic associations between data features and cognitive states. These limitations stem from LLMs' inability to reason about temporally and spatially structured data, as well as the lack of domain-specific causal reasoning mechanisms. This paper partially enhances these shortcomings by adding semantic information to eye-tracking data through a multi-stage data segmentation process. The highest accuracy can reach 50\% in difficulty prediction tasks.

However, this approach still relies on carefully designed prompts and prior knowledge, which may affect the generalization ability and adaptability of the model in some cases. In addition, LLMs' sensitivity to input may also cause the model to be unstable when processing complex or noisy data. Therefore, future research can explore combining LLMs with other perceptual models to form a rigid multimodal learning framework to improve the performance and robustness of the system when processing non-text data.

\subsection{Limitations}
While our results are promising, several limitations should be considered. First, our framework was primarily evaluated in the context of programming tasks using a specific educational dataset. Its generalizability to other cognitive domains, such as reading comprehension or creative reasoning, still needs empirical validation. Second, while the co-evaluation mechanism incorporates citation counts and journal quality, the weighting lacks finer granularity and doesn’t fully account for contextual domain alignment. This can lead to inflated validity scores for studies with high citation popularity but lower relevance. Enhancing this rubric and integrating more advanced retrieval-augmented generation (RAG) tools could improve future iterations. 

Additionally, though structured prompting improved performance, LLMs sometimes relied on superficial features (e.g., problem numbering) in difficulty prediction, indicating they hadn't fully captured the cognitive complexity of tasks. This highlights the need for better prompt engineering, stronger grounding in cognitive theory, and the potential integration of neuro-symbolic reasoning techniques. Lastly, in the anomaly detection module, while experts typically outperform students, there remains the possibility of student outperformance on specific tasks, which may occasionally lead to misclassifications. Finally, the absence of weighting factors for citation count and relevance score in calculating the final score F could introduce minor inconsistencies, which may slightly affect reliability but doesn’t significantly impact the overall results.

\section{Safe and Responsible Innovation Statement}
Our framework aims to support more accurate and interpretable analysis of cognitive states using eye-tracking data, but it also raises important ethical questions. When combined with large language models, cognitive modeling can unintentionally reveal sensitive information about individuals, even from anonymized data. This is especially critical in settings like education or healthcare, where privacy, consent, and fairness must be carefully protected.To address these concerns, we use only publicly available datasets and follow strict data usage policies. We also avoid any attempts to infer personal identities or make high-stakes decisions without human input. A key part of our approach is a human–AI co-evaluation mechanism that helps verify model outputs through expert review and literature support, making the system’s reasoning more transparent and trustworthy. While we take steps to reduce bias, such as using diverse models and prompt types, some risks remain. For instance, a student’s unusual but valid strategy could be flagged as an error, potentially leading to misleading behavioral patterns. We see this as an area for future work, especially in developing mitigation techniques that help the system better account for diverse behaviors and minimize unfair judgments
\vspace{-.20cm}

\section{Conclusion}
In this paper, we present a human–AI collaborative framework that enhances the reliability and interpretability of eye-tracking data analysis. By combining multi-stage segmentation, expert–LLM co-evaluation, and LSTM–LLM–based anomaly detection, the framework enables robust detection of attention shifts and cognitive states. Our findings show strong performance across LLMs and prompt strategies, with up to 50\% accuracy improvement in programming task assessments. This work offers a scalable, interpretable approach for cognitive state analysis and educational feedback, with broad implications for user-aware intelligent systems.
\bibliographystyle{ACM-Reference-Format}
\bibliography{sample-base}

\end{document}